\pdfoutput=1
\documentclass{PoS}

\usepackage{amsmath}

\def\la{\langle}
\def\ra{\rangle}

\def\ol{\overline}

\def\beq{\begin{equation}}
\def\eeq{\end{equation}}
\def\bea{\begin{eqnarray}}
\def\eea{\end{eqnarray}}
\def\barr{\begin{array}}
\def\earr{\end{array}}

\def\op{{\mathcal{O}}}

\title{Moments of pion distribution amplitude using operator product
  expansion on the lattice}

\ShortTitle{Pion distribution amplitude and lattice OPE}

\author{William Detmold\\
        Centre for Theoretical Physics, Massachusetts Institute of
        Technology, MA 02139, USA\\
        E-mail: \email{wdetmold@mit.edu}}

\author{Issaku Kanamori\\
        Department of Physical Science, Hiroshima University, Higashi-Hiroshima
        739-8526, Japan\\
        E-mail: \email{kanamori@hiroshima-u.ac.jp
}}

\author{C.-J.~David~Lin\\
        Institute of Physics, National Chiao-Tung University,
        Hsinchu 30010, Taiwan\\
        Centre for High Energy Physics, Chung-Yuan Christian
        University, Chung-Li, 32032, Taiwan
        E-mail: \email{dlin@mail.nctu.edu.tw}}

\author{\speaker{Santanu Mondal}\\
        Institute of Physics, National Chiao-Tung University,
        Hsinchu 30010, Taiwan\\
        E-mail: \email{santanu.sinp@gmail.com}}

\author{Yong Zhao\\
        Centre for Theoretical Physics, Massachusetts Institute of
        Technology, MA 02139, USA\\
        E-mail: \email{yzhaoqcd@mit.edu}}

\abstract{We report an exploratory study of the
current-current matrix elements that are relevant to the extraction of
moments of the pion light-cone distribution amplitude, employing the
method of introducing a valence relativistic heavy quark. 
The numerical investigation is carried out in the quenched approximation with
the physical volume $L\approx 2.4$ fm at two values of lattice spacing (0.05
and 0.075 fm).  We
obtain clean signals for the relevant Euclidean hadronic tensor with
reasonable statistics, but observe 
that the lattice artefacts are non-negligible in our results.  The key conclusion from the analysis
hitherto is that although our approach has the potential for making significant
contributions to parton physics, data at finer lattice
spacings that are currently being produced are needed in order to control the continuum extrapolation.}

\FullConference{The 36th Annual International Symposium on Lattice Field Theory - LATTICE2018\\
		22-28 July, 2018\\
		Michigan State University, East Lansing, Michigan, USA.}

\begin{document}

\section{Introduction}
\label{sec:introduction}
Numerical implementation of lattice field theory is normally carried out in
Euclidean space.  Therefore it is challenging to apply lattice QCD to
extract reliable results in parton physics, which involves non-perturbative dynamics on
the light cone.  For this reason, the traditional approach to extracting various parton
distribution functions (PDF's) and light-cone distribution amplitudes
(LCDA's) employing lattice QCD relies on the calculation of Mellin moments.
These moments are related to matrix elements of local operators that arise from an
operator product expansion (OPE).   Because O(4) Euclidean space-time
symmetry is broken by the lattice geometry, these local operators mix under
renormalisation, resulting in power divergences that
are difficult to subtract accurately~\cite{Kronfeld:1984zv,Martinelli:1987si}.  This
is the reason why lattice-QCD determination of the
PDF's and the LCDA's using the above strategy has only been giving results for the first three Mellin moments.

Alternative methods for gaining parton-physics information with
lattice QCD have been suggested over the past two
decades~\cite{Aglietti:1998ur,Liu:1999ak,Detmold:2005gg,Braun:2007wv,Davoudi:2012ya,Ji:2013dva,Chambers:2017dov,Radyushkin:2017cyf,Ma:2017pxb}.
These methods involve calculating hadronic matrix elements of non-local
operators, and many of them are presently under intensive investigation~\cite{Monahan:latt18}.
In this article, we present progress of performing a lattice
calculation for the pion LCDA, employing the proposal of introducing a
valence relativistic heavy quark, as detailed for the quark PDF's in
Ref.~\cite{Detmold:2005gg}.  
The pion LCDA, $\phi_{\pi} (\xi)$, is of importance in understanding hadronic exclusive
decay processes in QCD~\cite{Lepage:1980fj}, as well as in extracting information in
flavour physics~\cite{Beneke:1999br}.  It is defined as
\beq
\label{eq:pion_DA_def}
\la 0 | \bar{d}(-z) \gamma_{\mu} \gamma_{5} {\mathcal{W}}[-z, z] u(z) |
\pi^{+} (p) \ra =
 i  p_{\mu} f_{\pi} \int_{-1}^{1} d \xi \mbox{ }
 {\mathrm{e}}^{-i \xi p\cdot z }\phi_{\pi}(\xi) ,
\eeq
with $z^{2}=0$, and ${\mathcal{W}}[-z, z]$ being a light-like
Wilson line between $-z$ and $z$.  The variable $\xi$ represents the
fraction of the pion momentum carried by the valence $u$ quark.  The
above DA can be constructed from its Mellin moments, $a_{n}$,  that
are related to local matrix elements in QCD,
\bea
\label{eq:moments_def}
&& a_{n} = \int_{0}^{1} d\xi \mbox{ }\xi^{n} \mbox{ } \phi_{\pi}(\xi)
\, , 
%\mbox{ }{\mathrm{and}}
\nonumber\\
&& f_{\pi}~a_{n-1}~[p^{\mu_1}\dots p^{\mu_n}-{\rm
  Traces}]= -i \langle 0|\bar{d}\gamma^{\{ \mu_1}\gamma^5(iD^{\mu_2})\dots
(iD^{\mu_n\}}) u
-{\rm Traces} \arrowvert
\pi ^+(p)\rangle ,
\eea
with the Lorentz indices symmetrised.  From
early~\cite{Martinelli:1987si} to
recent~\cite{Braun:2006dg, Arthur:2010xf, Braun:2015axa, Bali:2017ude} lattice
calculations following the traditional approach, only the second
moment of this LCDA has been extracted because of the above issue of power
divergence in the operator mixing.   
Using the strategies in Refs.~\cite{Braun:2007wv,Ji:2013dva}, exploratory
results for the $\xi{-}$dependence of $\phi_{\pi}(\xi)$ have recently been reported~\cite{Zhang:2017bzy,Bali:2018spj}.

\section{Operator product expansion and the valence heavy quark}
\label{sec:OPE_and_HQ}
Using the method of
Ref.~\cite{Detmold:2005gg}, it can be shown that
the Euclidean hadronic tensor,
\beq
\label{eq:hadronic_tensor}
  U^{ [\mu\nu]}_{A} (q,p) 
=\int ~d^4x ~ e^{iqx}~\langle 0|T[A^{[\mu}_{\Psi,\psi}(x)~
A^{\nu]}_{\Psi,\psi}(0)]\arrowvert\pi^+(p)\rangle ,
\eeq
in the continuum limit enables one to extract the moments, $a_{n}$,
defined in Eq.~(\ref{eq:moments_def}) without having to subtract any
power divergence.  In Eq.~(\ref{eq:hadronic_tensor}),
the Lorentz indices, $\mu$ and $\nu$, are antisymmetrised, and
the axial current is defined as
\beq
\label{eq:axial_current}
A^\mu_{\Psi,\psi}=\ol{\Psi}\gamma^\mu\gamma_5\psi+\ol{\psi}\gamma^\mu\gamma_5\Psi ,
\eeq
with $\psi$ being a light-quark and $\Psi$ being the
valence heavy-quark fields.  We stress that the approach outlined in
Ref.~\cite{Detmold:2005gg} requires the extrapolation of lattice
results for $U^{ [\mu\nu]}_{A} (q,p)$ to the continuum limit.
Furthermore, the hadronic tensor, 
$U^{ [\mu\nu]}_{A} (q,p)$, should be computed in the ``unphysical'' regime
\beq
\label{eq:unphysical_regime}
 (p_{M}+q_{M})^{2} < (m_{\Psi} + \Lambda_{{\mathrm{QCD}}})^{2} ,
\eeq
where $q_{M}$ and $p_{M}$ are the
Minkowskian counterparts of $q$ and $p$.
%where there are no multi-particle on-shell states between the two axial
%currents in Eq.~(\ref{eq:hadronic_tensor}).   
In this regime the analytic continuation of $U^{ [\mu\nu]}_{A} (q,p)$ to Minkowski space is
straightforward.  It can be achieved by simply relating
$q_{4}$ to $iq_{0}$.
The above
considerations lead to the 
requirement for the hierarchy of scales
\beq
\label{eq:scale_hierarchy}
 \Lambda_{{\mathrm{QCD}}} \ll \sqrt{q^{2}} \lessapprox m_{\Psi} \ll \frac{1}{a},
\eeq
where $a$ is the lattice spacing.

Performing an OPE by following the procedure in Ref.~\cite{Detmold:2005gg}, one obtains
\beq
\label{eq:ope}
U_A^{[\mu\nu]}(p,q)
 =   2 i f_\pi \epsilon_{\mu\nu\rho\lambda}
 q^{\rho}p^{\lambda}\times
\sum_{n=0,2,4\cdots}^{\infty}
 \left [
\frac{\zeta^{n}C^2_n(\eta)}{(n+1) \tilde{Q}^2}
\right ]
 {\mathcal{C}}_{W}^{(n)} \left ( m_{\Psi}, \tilde{Q}, \mu \right ) a_n (\mu) , 
\eeq
where 
%$\zeta=\frac{\sqrt{p^2q^2}}{\tilde{Q}^2},
$\zeta=\sqrt{p^2q^2}/\tilde{Q}^2$,
%$\eta=\frac{p.q}{\sqrt{p^2q^2}}$, 
$\eta=p.q/\sqrt{p^2q^2}$, 
$\tilde{Q}^{2} = -q^{2} - m_{\Psi}^{2} $, 
the ${\mathcal{C}}_{W}^{(n)}( m_{\Psi}, \tilde{Q}, \mu )$
are the Wilson coefficients, $\mu$ is the renormalisation scale, 
and the $C^2_n(\eta)$ are the Gegenbauer
polynomials that arise from resumming the target-mass effects.  Notice
that there is an ambiguity of $O(\Lambda_{{\mathrm{QCD}}})$ in the
definition of $\tilde{Q}$. Detailed discussion of this ambiguity and
the higher-twist contributions can
be found in Ref.~\cite{Detmold:2005gg}.
Also, Eq.~(\ref{eq:ope}) indicates that the hadronic tensor,
$U_A^{[\mu\nu]}(p,q)$, is purely imaginary in Euclidean space.

\section{Correlators and simulation details}
\label{sec:corr_and_numerical}
We consider the following correlators involving the pion interpolating
operator, $\op_{\pi}$, and the current in Eq.~(\ref{eq:axial_current}),
\bea
\label{eq:correlators}
 C_{3}^{\mu\nu} \left ( \tau_{e}, \tau_{s}; \vec{p}_{e}, \vec{p}_{s} \right ) &=&
 \int d^{3}x_{e}d^{3}x_{s} 
 {\mathrm{e}}^{i \vec{p}_{e}\cdot \vec{x}_{e} + i \vec{p}_{s}\cdot
   \vec{x}_{s}  }
\left \langle 0 \left | {\mathrm{T}} \left [ A^{\mu}_{\Psi,\psi} \left ( \vec{x}_{e}, \tau_{e}
   \right ) A^{\nu}_{\Psi,\psi} \left ( \vec{x}_{m}, \tau_{s} \right ) \op_{\pi}^{\dagger} (0)
 \right ] \right | 0 \right \rangle ,
\nonumber\\
C_{\pi} \left ( \tau_{\pi}; \vec{p} \right ) &=& \int
 d^{3}x \mbox{ } {\mathrm{e}}^{i \vec{p}\cdot \vec{x}}
  \left \langle 0 \left | \op_{\pi}(\vec{x},\tau_{\pi})
 \op_{\pi}^{\dagger}(0) \right  | 0 \right \rangle ,
\eea
where the subscripts, $e$ and $s$, in the three-point function stand
for ``extended'' and ``sink'' points in the computation of the 
quark propagators.  It is straightforward to demonstrate that
using $\mathrm{C_{3}^{\mu\nu}}$
and $\mathrm{C_{\pi}}$ in the limit where $\tau_{\pi,e,s}$ are all large, one can extract the quantity, $R_{3}^{\mu\nu}$, that is
defined as
\beq
\label{eq:R3_def}
R_3^{\mu\nu}(\tau,\vec{q},\vec{p})\equiv \int ~d^3x ~ e^{i\vec{q}. \vec{x}}~\langle
0|T[A^\mu_{\Psi,\psi}(\vec{x},\tau)~
A^\nu_{\Psi,\psi}(\vec{0},0)]\arrowvert\pi (p)\rangle.
\eeq
%
%The quantity $\mathrm{R_3^{\mu\nu}(\tau,\vec{q},\vec{p})}$ is the lattice version of
%$\mathrm{\int ~d^3x ~ e^{i\vec{q}. \vec{x}}~\langle
%0|T[A^\mu_{\Psi,\psi}(\vec{x},\tau)~
%A^\nu_{\Psi,\psi}(\vec{0},0)]\arrowvert\pi (p)\rangle}$. 
Thus the hadronic tensor in Eq.~(\ref{eq:hadronic_tensor}) can be obtained by 
performing the Fourier transform of $R_3^{[\mu\nu]}(\tau,\vec{q},\vec{p})$
in the temporal direction,
\beq
\label{eq:FT}
{U}^{[\mu\nu]}_A(q,p)=\int_{\tau_{{\mathrm{min}}}}^{\tau_{{\mathrm{max}}}} \mbox{ } d\tau ~ e^{iq_4\tau}
~R_3^{[\mu\nu]}(\tau,\vec{q},\vec{p}) ,
\eeq
where the integration range, $[\tau_{{\mathrm{min}}},
\tau_{{\mathrm{max}}}]$, is constrained by the range of Euclidean
time where we are able to isolate the one-pion state in $\mathrm{C_{3}^{\mu\nu}}$
and $\mathrm{C_{\pi}}$.  Because of the use of
the valence heavy quark, which exponentially localises the signal, we find that the above Fourier transform is
well approximated when $\tau_{{\mathrm{min}}}
\sim -0.7$ fm and $\tau_{{\mathrm{max}}} \sim 0.7$ fm for the
results reported here.  
%Therefore in this work we fix $\tau_{{\mathrm{min}}} \sim -0.75$ fm and allow
%$\tau_{{\mathrm{max}}}$ to vary between $\tau_{{\mathrm{min}}}$ and
%$\sim 0.75$ fm 
%(see plots in Figs.~\ref{fig1},~\ref{fig2} and \ref{fig3}).

Our simulations are performed in the quenched approximation with Wilson
gauge action.  Gauge field ensembles are generated at several values of
lattice spacing, $a$, and the finest lattice is at $a^{-1} = 8$ GeV\footnote{
Ensembles with $a^{-1} > 2.7$ GeV are generated by employing the
method proposed in Ref.~\cite{Endres:2015yca}.}.   All these ensembles
are tuned to have the same spatial lattice size, $L \approx 2.4$ fm,
with the temporal extent being $2L$.  Valence fermion propagators
are computed employing 
non-perturbatively $O(a)$-improved Wilson action, with values of the clover
coefficient determined using the result of
Ref.~\cite{Edwards:1997nh}.   
Since this is an exploratory
investigation, we use the one-loop matching coefficient, $Z_{A}$, for
the axial current  in
Eq.~(\ref{eq:axial_current})~\cite{Gabrielli:1990us}, without $O(a){-}$improving it.  However, the quark-field normalisation
proposed in Ref.~\cite{ElKhadra:1996mp} is employed to reduce lattice artefacts.

At this conference, our presentation focuses on calculations from two lattice
spacings, $a = 0.075$ fm ($a^{-1} = 2.67$ GeV) and $a = 0.05$ fm ($a^{-1} = 4$ GeV), and at two choices
of heavy-quark masses, $m_{\Psi} = 1.3$ GeV and 2 GeV\footnote{We also
have exploratory results at $a = 0.06$ fm ($a^{-1} = 3.33$ GeV) for
$m_{\Psi} = 1.3$ GeV.  They are included in Fig.~\ref{fig:had_tensor_p000_3_lat}.}.  
The input bare light-quark mass is tuned such that the pion mass $M_\pi = 450$
MeV in this exploratory study.

\section{Exploratory results}
\label{sec:results}
\begin{figure}[t!]
\begin{center}
\vspace{-0.3cm}
        \includegraphics[width=7.5cm,height=6.0cm]{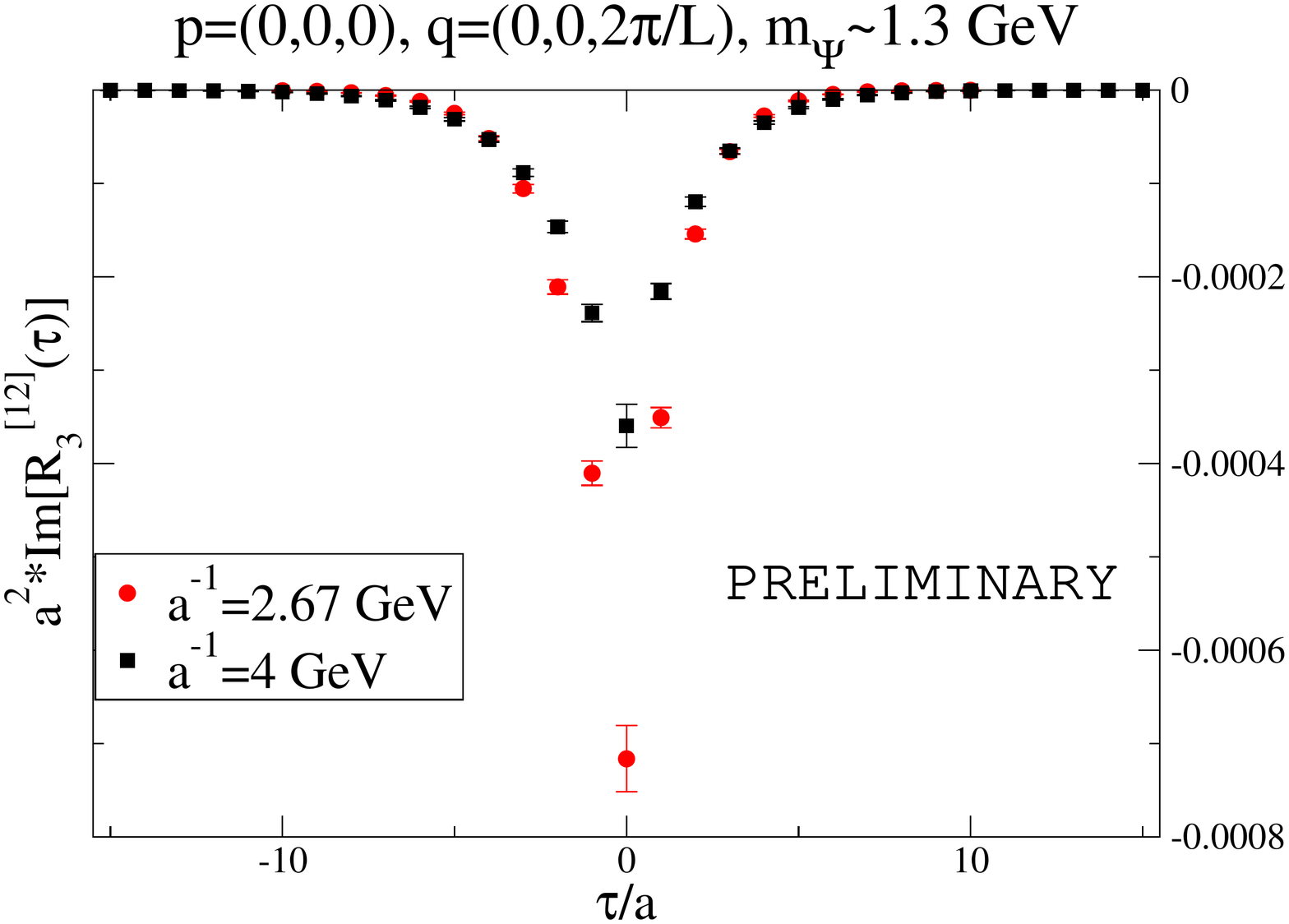}
    \hspace{-0.2cm}
        \includegraphics[width=7.5cm,height=6.0cm]{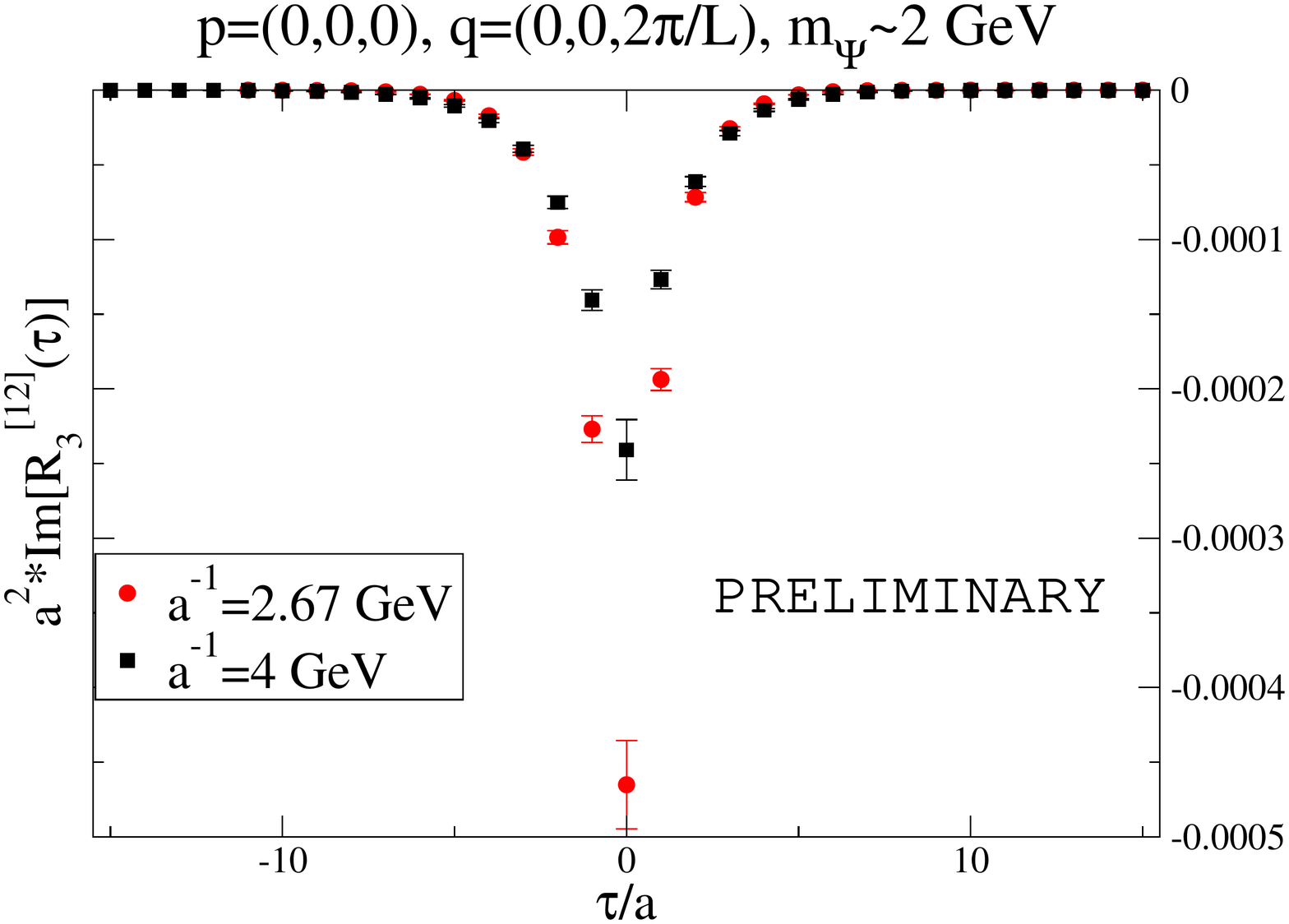}
\vspace{-0.5cm}
       \caption{Imaginary part of bare $R_{3}^{[12]}(\tau,\vec{q},\vec{p})$ with
         $\vec{q}=(0,0,2\pi/L)$ and $\vec{p}=(0,0,0)$, at $m_\Psi=1.3$ GeV
         (left) and $m_\Psi=2$ GeV (right). }
\label{fig:R3_p000}
\end{center}
\end{figure}
\begin{figure}[t!]
\begin{center}
\vspace{-0.3cm}
         \includegraphics[width=7.5cm,height=6.0cm]{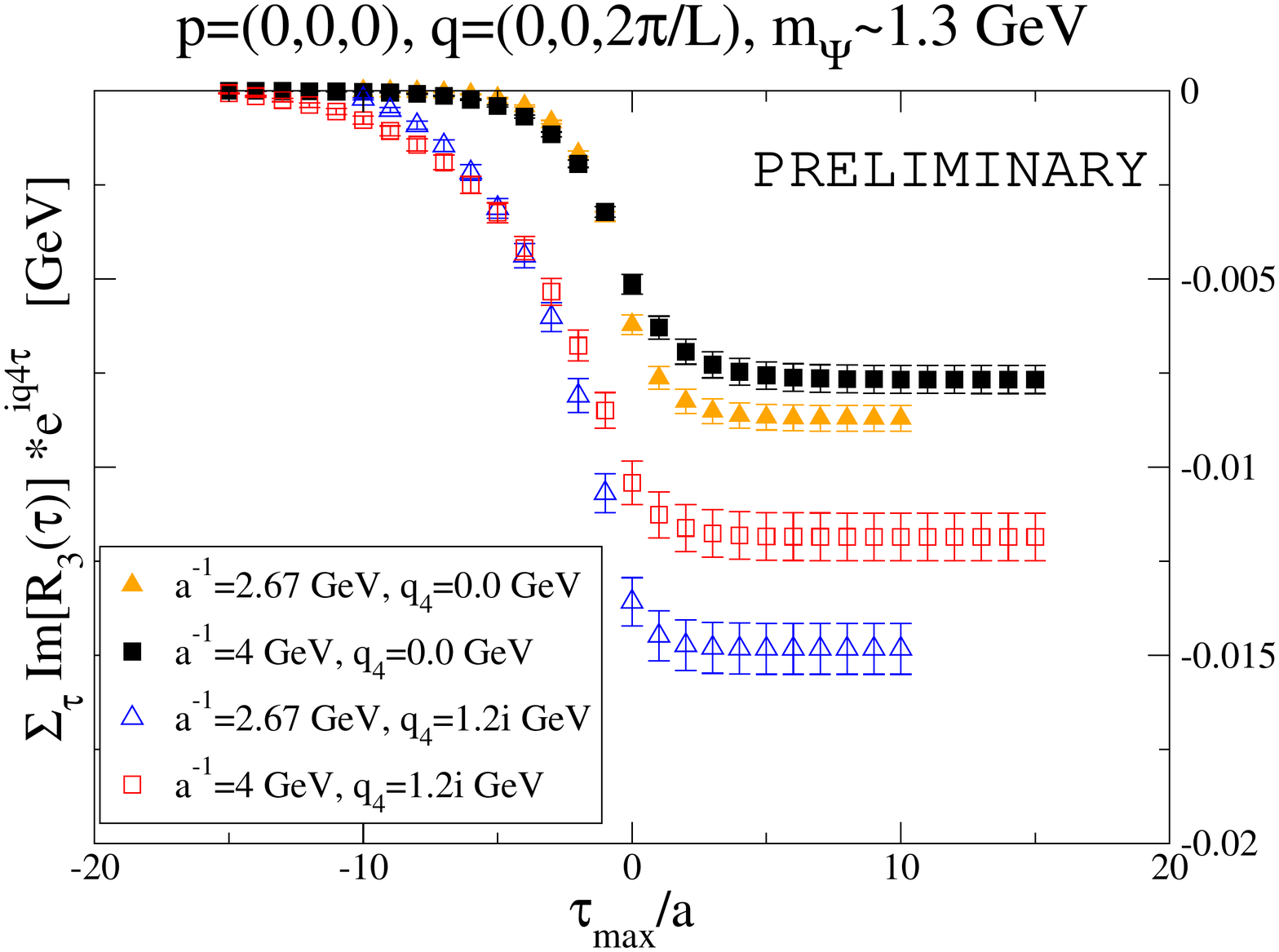}
    \hspace{-0.2cm}
        \includegraphics[width=7.5cm,height=6.0cm]{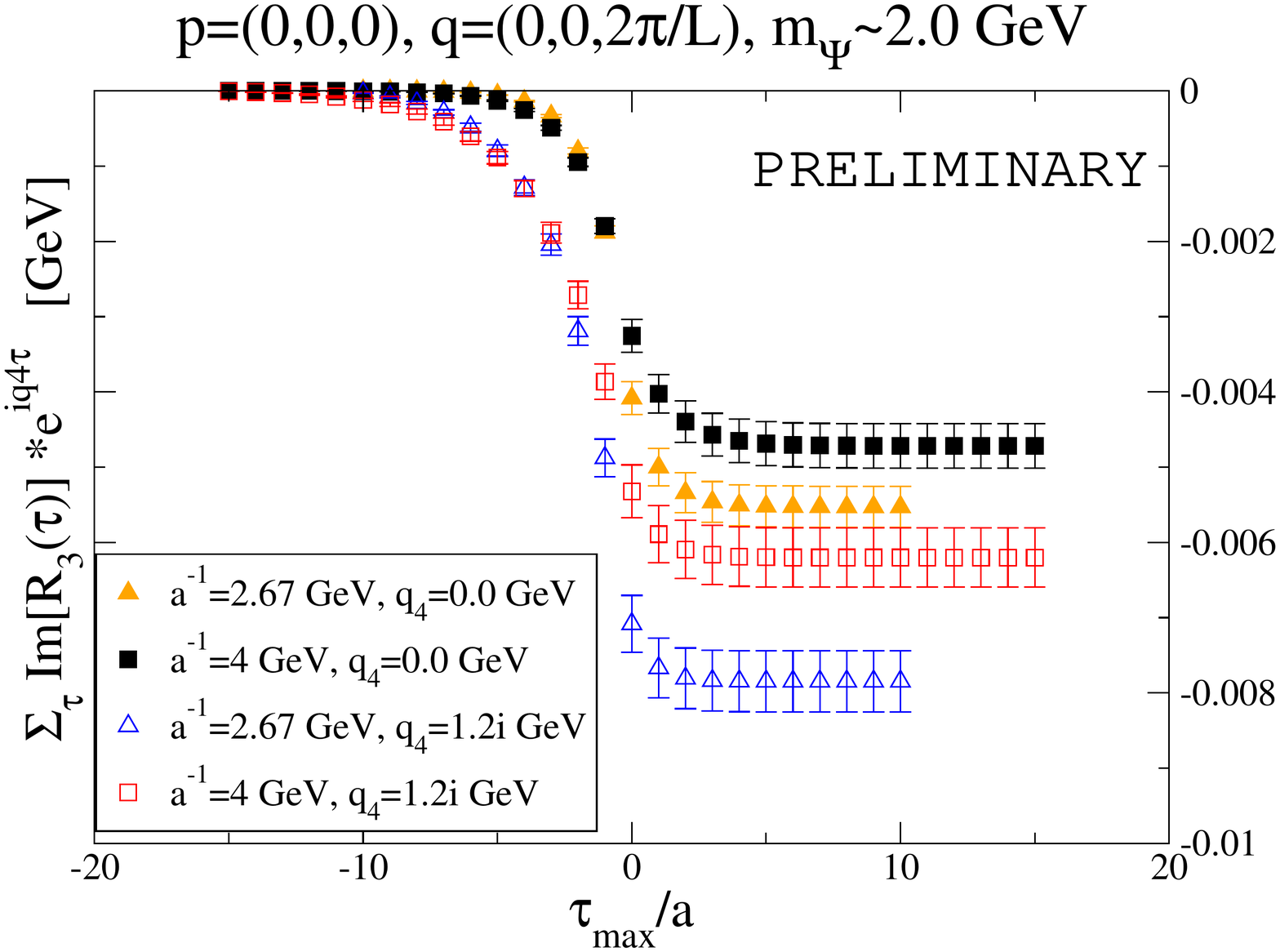}
\vspace{-0.5cm}
       \caption{Time-direction Fourier transform for obtaining bare
         Im$[U_{A}^{[12]}(q,p)]$ with
         $\vec{q}=(0,0,2\pi/L)$ and $\vec{p}=(0,0,0)$ and two choices
         $q_{4}$, at $m_\Psi=1.3$ GeV (left) and 2 GeV
         (right).}
\label{fig:FT_p000}
\end{center}
\end{figure}
\begin{figure}[t!]
\begin{center}
\vspace{-0.3cm}
        \includegraphics[width=8.5cm,height=6.5cm]{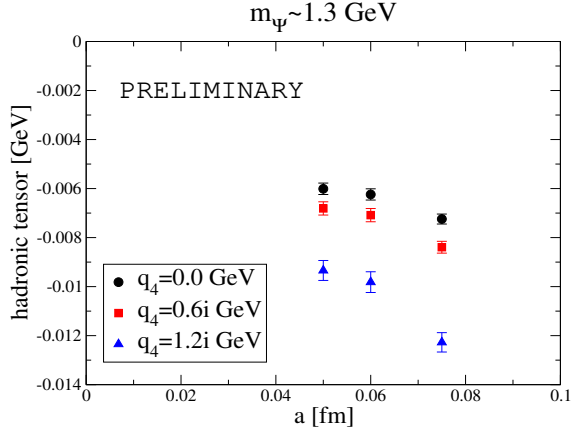}
\vspace{-0.5cm}
       \caption{Imaginary part of $U_{A}^{[12]}(q,p)$ for
         $\vec{q}=(0,0,2\pi/L)$, $\vec{p}=(0,0,0)$ and $m_\Psi=1.3$ GeV,
         at $a=0.05$, 0.06, 0.075 fm ($a^{-1}=4.0$, 3.33, 2.67 GeV, respectively), with 3 choices of $q_{4}$.   One-loop $Z_{A}$ is used.}
\label{fig:had_tensor_p000_3_lat}
\end{center}
\end{figure}

Figure~\ref{fig:R3_p000} shows results for the imaginary part of bare 
$R_{3}^{[12]}(\tau,\vec{q},\vec{p})$ with $\vec{q}=(0,0,2\pi/L)$ and
$\vec{p}=(0,0,0)$, at $m_\Psi=1.3$ GeV and 2 GeV.
The temporal-direction Fourier transforms [Eq.~(\ref{eq:FT})] on these ratios for obtaining
the corresponding hadronic tensors are displayed in
Fig.~\ref{fig:FT_p000}, where we are showing the cases of $q_{4}=0$
and $q_{4}=1.2 i$ GeV.
For this choice of the momenta, result of the OPE for the hadronic
tensor is significantly simplified, and in principle
allows us to extract the moments, $\{ a_{n} \}$, through varying
$q_{4}$.  However, we do not attempt such numerical exercise at this stage,
because the lattice artefacts are observed to be non-negligible, as
evidenced in Figs.~\ref{fig:FT_p000} and
\ref{fig:had_tensor_p000_3_lat}.   Notice that we also include the
exploratory results from the $a=0.06$ fm lattice in
Fig.~\ref{fig:had_tensor_p000_3_lat}.   As stressed earlier in this
article, our strategy requires reliable extrapolation of the hadronic tensor to the continuum
limit.

\begin{figure}[t!]
\begin{center}
\vspace{-0.3cm}
        \includegraphics[width=7.5cm,height=6.0cm]{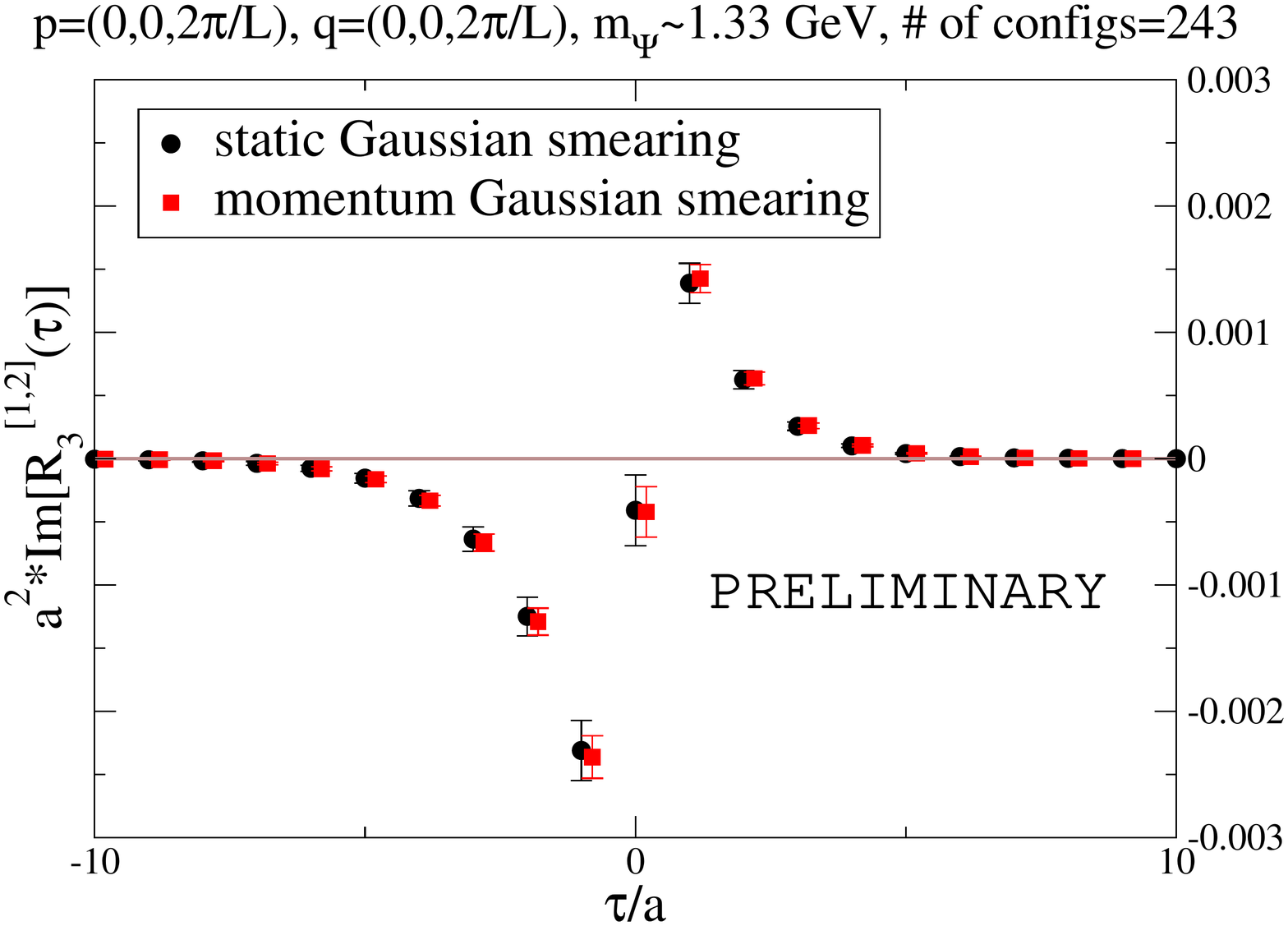}
    \hspace{-0.2cm}
        \includegraphics[width=7.5cm,height=6.0cm]{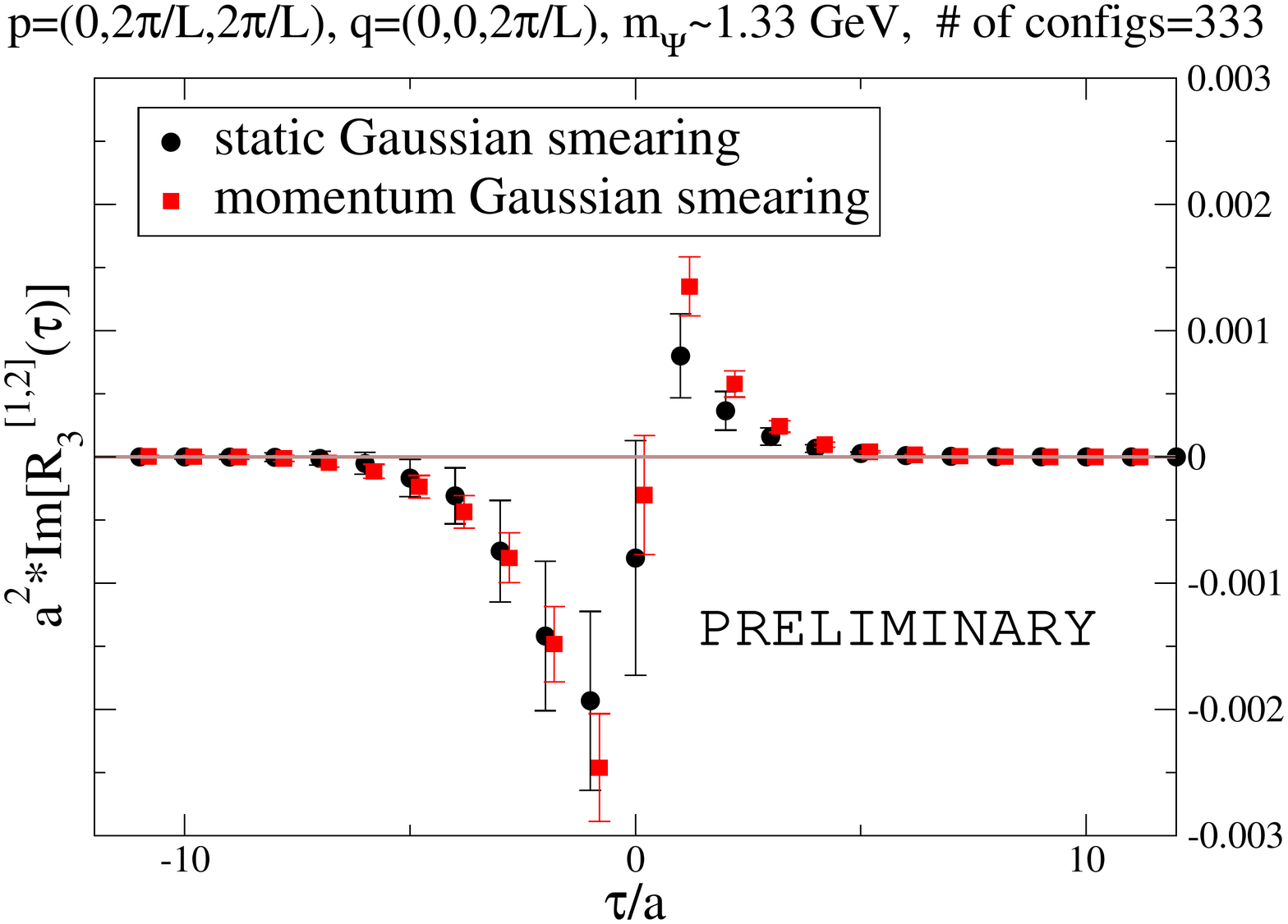}
\vspace{-0.5cm}
       \caption{Bare $R_{3}^{[12]}(\tau,\vec{q},\vec{p})$ with
         $\vec{q}=(0,0,2\pi/L)$ and  $m_\Psi=1.3$ GeV at $\vec{p}=(0,0,2\pi/L)$
         (left) and $\vec{p}=(0,2\pi/L,2\pi/L)$ (right).  Black circle is for the usual gauge invariant Gaussian source and sink,
red square is for the momentum smearing with momentum $\pm 0.7\vec{p}$
for quark and anti-quark.}
\label{fig:R3_p001_and_011}
\end{center}
\end{figure}

\begin{figure}[t!]
\begin{center}
\vspace{-0.3cm}
         \includegraphics[width=7.5cm,height=6.0cm]{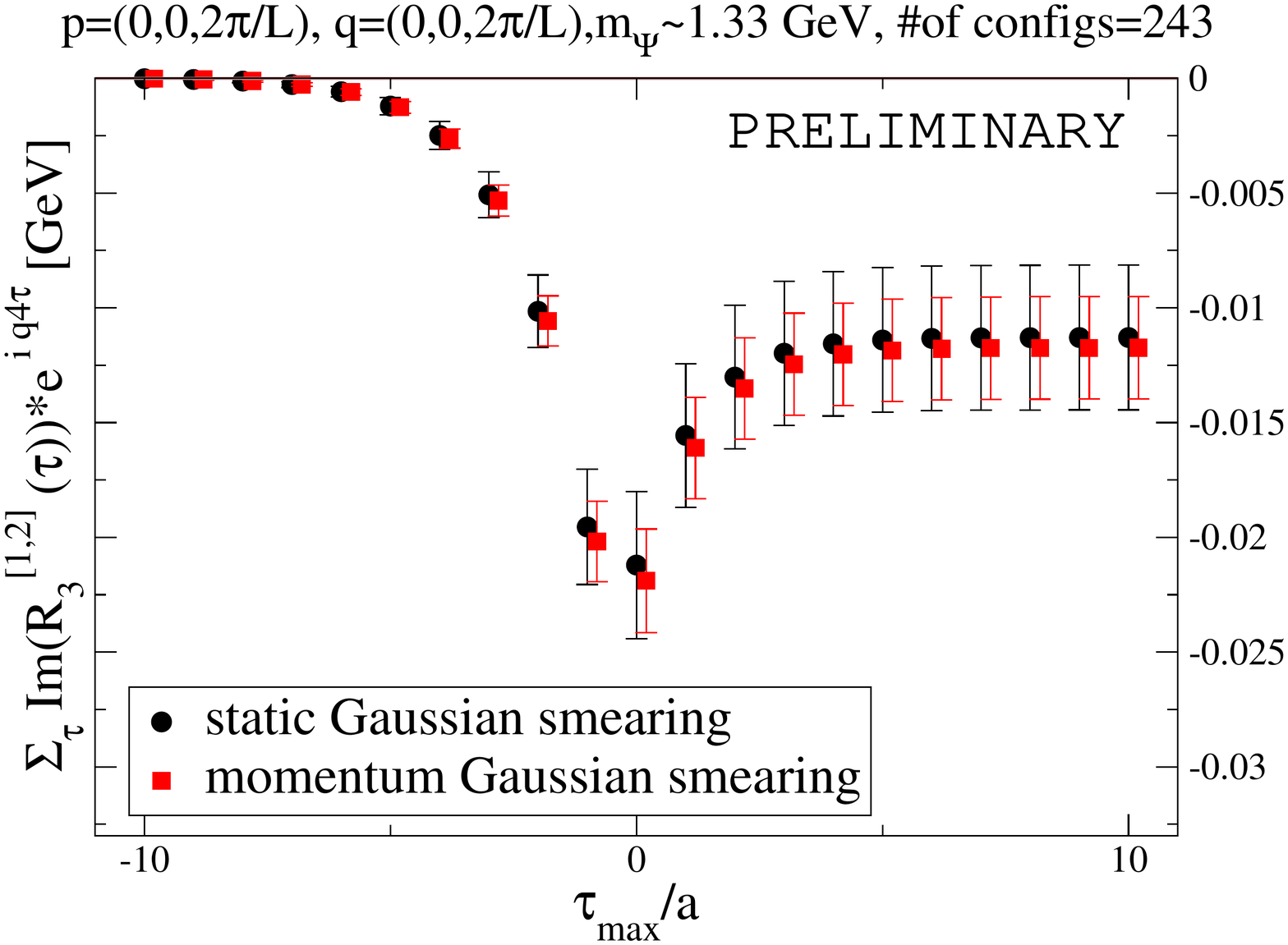}
    \hspace{-0.2cm}
        \includegraphics[width=7.5cm,height=6.0cm]{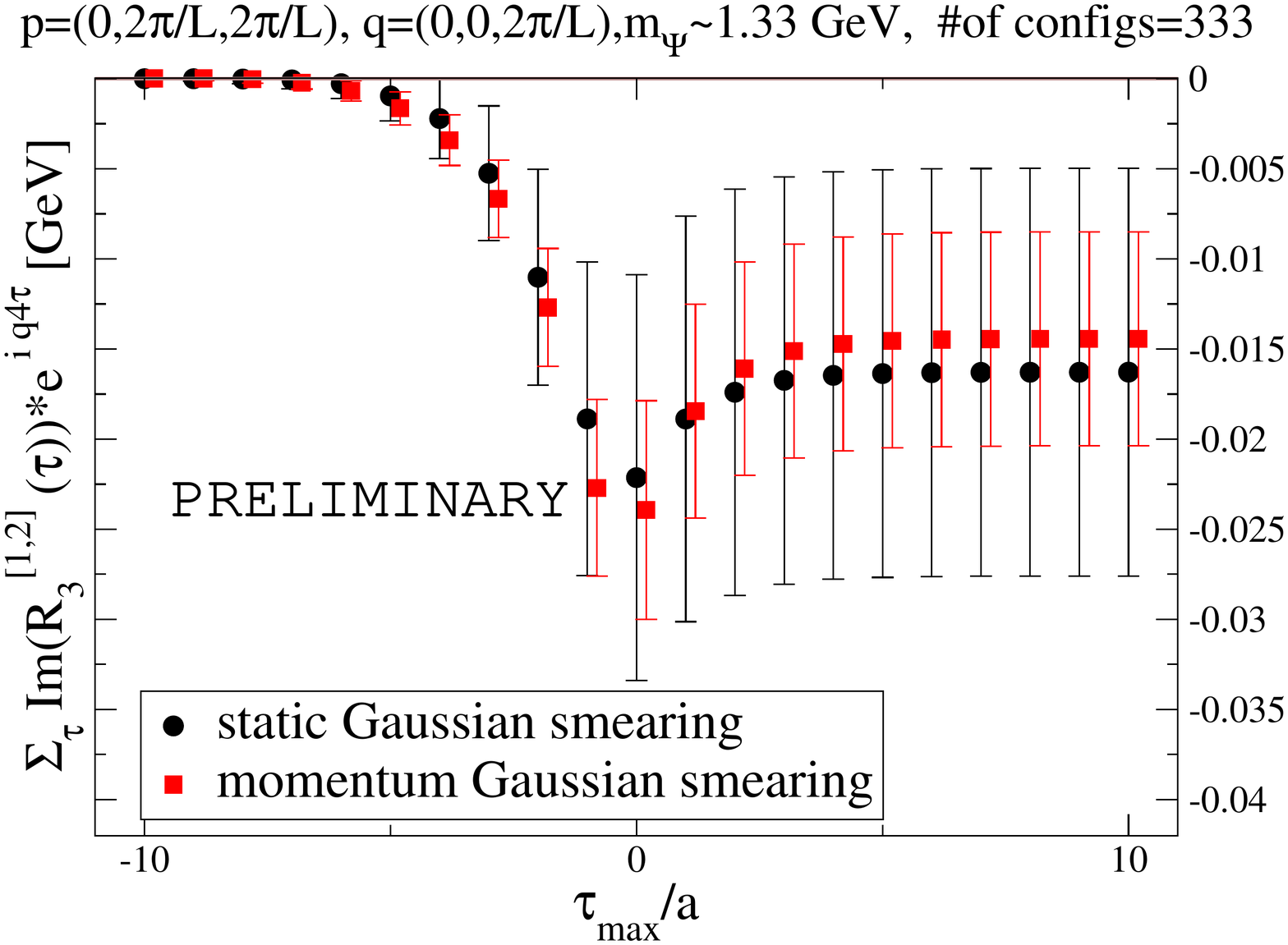}
\vspace{-0.5cm}
       \caption{Bare Im$[U_{A}^{[12]}(q,p)]$with
         $\vec{q}=(0,0,2\pi/L)$ and  $m_\Psi=1.3$ GeV at $\vec{p}=(0,0,2\pi/L)$
         (left) and $\vec{p}=(0,2\pi/L,2\pi/L)$ (right).  Black circle is for the usual gauge invariant Gaussian source and sink,
red square is for the momentum smearing with momentum $\pm 0.7\vec{p}$
for quark and anti-quark.}
\label{fig:FT_p001_and_011}
\end{center}
\end{figure}

We have also examined $R^{[12]}_{3}(\tau,\vec{q},\vec{p})$ and
$U^{[12]}_{A}(p,q)$ at non-vanishing pion momentum.  In this case, the
lattice data are expected to be noisy.  To address this, we have
investigated the technique of momentum smearing~\cite{Bali:2016lva} for the pion
interpolating operator, ${\mathcal{O}}_{\pi}$.  Results of the study
for two choices of the pion momentum, $\vec{p} = (0,0,2\pi/L)$ and 
$\vec{p} = (0, 2\pi/L, 2\pi/L)$, at the current-injected momentum
$\vec{q}=(0,0,2\pi/L)$ 
are presented in Figs.~\ref{fig:R3_p001_and_011} and \ref{fig:FT_p001_and_011}.
Plots in these figures show that  momentum smearing is
advantageous already for these low values of $|\vec{p}|$, although its
implementation requires
separate computations for the light quark and anti-quark propagators.
We will make use of this technique in our future work on the pion LCDA.

\section{Conclusion and outlook}
\label{sec:outlook}
In this article, we report progress of our exploratory investigation
of the pion LCDA using the strategy of introducing a valence heavy
quark in the current-current correlator. This strategy allows us to extract higher moments
for this LCDA.  We find that reasonably good signals can be obtained for the
relevant Euclidean hadronic tensor.  Although this demonstrates a
promising future of this approach, we observe that lattice
artefacts can still be non-negligible in the regime $0.05\mbox{
}{\mathrm{fm}}\lessapprox a\lessapprox 0.075$ fm.  Since our method requires reliable continuum
extrapolation, it is necessary to have data at finer lattice spacings
for this task, as will be studied in the near future.

\section*{Acknowledgments}
We grateful 
to Michael Endres for generating the gauge configurations, and to Balint Joo for helping us with the QPhiX library.  WD and YZ
are supported by the U.S. Department of Energy, Office of Science,
Office of Nuclear Physics, from de-sc0011090 and within the framework
of the TMD Topical Collaboration, and thank National Chiao-Tung
University for hospitality.
WD is also supported by the U.S. Department of Energy under Early Career Research Award de-sc0010495 and the SciDAC4 award de-sc0018121. IK acknowledges Priority Issue 9 to be tacked by Using Post K Computer
and Joint Institute for Computational Science (JICFuS).  CJDL and SM are supported by Taiwanese MoST via grants
105-2112-M-002-023-MY3, and thank MIT for hospitality.   We acknowledge support by the MIT MISTI program.

\end{document}